\documentclass[pre,preprint,showpacs,showkeys,preprintnumbers,amsmath,amssymb]{revtex4} 
 
\usepackage{graphicx}

\usepackage{dcolumn}

\usepackage{bm}

\begin{document}

\newcommand{\be}{\begin{equation}}  
\newcommand{\ee}{\end{equation}}
\newcommand{\mean}[1]{\left\langle #1 \right\rangle}
\newcommand{\abs}[1]{\left| #1 \right|}
\newcommand{\set}[1]{\left\{ #1 \right\}}
\newcommand{\norm}[1]{\left\|#1\right\|}
\newcommand{\eps}{\varepsilon}
\newcommand{\tNN}{\tilde{\mathbf{X}}_n^{NN}}
\newcommand{\NN}{\mathbf{X}_n^{NN}}
\newcommand{\ber}{\begin{eqnarray}}
\newcommand{\eer}{\end{eqnarray}}
\newcommand{\lb}{\left(}
\newcommand{\rb}{\right)}

\preprint{Chaos}

\title{Improvement of speech recognition by nonlinear noise reduction}

\author{Krzysztof Urbanowicz}
 \email{urbanow@pks.mpg.de}
 \affiliation{Max Planck Institute for Physics of Complex Systems\\
 N\"{o}thnitzer Str. 38\\ D--01187 Dresden, Germany}
\affiliation{Faculty of Physics, Warsaw University of Technology\\
 Koszykowa 75\\ PL--00-662 Warsaw, Poland}
\author{Holger Kantz}
 \email{kantz@pks.mpg.de}
\affiliation{Max Planck
Institute for Physics of Complex Systems\\
 N\"{o}thnitzer Str. 38\\ D--01187 Dresden, Germany}
\date{\today}

\begin{abstract}
The success of nonlinear noise reduction applied to a single
channel recording of human voice is measured in terms of the
recognition rate of a commercial speech recognition program in
comparison to the optimal linear filter. The overall performance
of the nonlinear method is shown to be  superior. We hence
demonstrate that an algorithm which has its roots in the theory of
nonlinear deterministic dynamics possesses a large potential in a
realistic application.
\end{abstract}
 \maketitle {\bf It is a common nuisance
that in signal recording, signal transmission, and signal storage,
perturbations occur which distort the original signal. In many
cases these perturbations are purely additive, so that in
principle they could be removed once identified. If these
perturbations are sufficiently uncorrelated and appear to be
random, they are called noise. Noise reduction is hence an
important data processing task. Traditionally, the distinction between
signal and noise is done in the frequency domain, where noise
contributes with a broad background to a signal which should
occupy distinguished frequency bands only. If the signal itself is
irregular in time and hence has a broad power spectrum, spectral
methods have a poor performance. Nonlinear methods for noise
reduction assume and exploit constraints in the clean signal which
are beyond second order statistics and hence are most efficiently
captured by non-parametric dynamical modelling of the signal. In
this paper it is shown that a particular such algorithm is able to
denoise a recorded human voice signal so that the recognition rate
of a commercial speech recognition program is significantly
enhanced. Hence, we have not only a more suitable quantifier for
the success of noise reduction on human voice, but we also verify
that noise reduction algorithms do what they should do, namely
enhance the intelligibility of the signal. This latter aspect is
highly non-trivial, since every noise reduction algorithm when
removing noise also distorts the signal. Finding merely a positive
gain hence does not guarantee that a human (or algorithm) really
understands better the meaning of the signal.}
\section{Introduction}
Noise reduction and source separation are ubiquitous data analysis
and signal processing tasks. In the analysis of chaotic data,
noise reduction became a prominent issue about 15 years ago
\cite{Schreiber1,kostelich,Schreiber2,Farmer,hammel,davies,Hsu,Sauer,GHKSS}.
Since the analysis of chaotic data in terms of dimensions,
entropies and Lyapunov exponents requires an access to the small
length scales (small scale fluctuations of the signal), already a
moderate amount of measurement noise on data is known to be
destructive. On the other hand, a deterministic source of a
signal, albeit potentially chaotic, supplies redundancy which
enables one to distinguish between signal and noise and eventually
to remove the latter to some extend. Noise reduction schemes which
exploit such dynamical constraints were proposed in
\cite{Schreiber1,kostelich,Schreiber2,Farmer,hammel,davies,Hsu,Sauer,GHKSS}
and where tested on many data sets. Since such algorithms were
designed to treat chaotic data, they do not make use of spectral
properties of data and can, in principle, even remove in-band
noise, which is noise whose high frequency spectrum is identical
to the spectrum of the signal. \par Human voice is a typical
non-stationary signal, where noise reduction is a relevant issue.
In telecommunication, in the development of hearing aids, and in
automatic speech recognition, noise contamination of the speech
signal poses severe problems. In multiple simultaneous
recordings, noise reduction is also known as blind source
separation. However, most often a single recording only is
available. In a previous study \cite{kantzhuman,IEEE} we
demonstrated that nonlinear noise reduction can cope with noise on
human speech data and has a performance, which is comparable to
advanced spectral adaptive filter banks.
The main differences in concepts of linear and nonlinear methods of
noise reduction used in this paper are listed in Table~\ref{tab.1}.
\par The performance of a noise reduction scheme is usually
measured as gain in dB. For this purpose, one starts with a clean
signal $\tilde{x}(t)$, numerically adds noise to $\tilde{x}(t)$
resulting in $x(t)$, and then applies the noise reduction scheme
without making use of $\tilde{x}(t)$. When we call the result of
the noise reduction $y(t)$, then the gain is defined as the
logarithm of the ratio of the noise power before and after the
noise reduction, which is 
\be 
\mbox{gain} = 10\log_{10}
\left(\frac{\langle(x(t)-\tilde{x}(t))^2\rangle}
{\langle(y(t)-\tilde{x}(t))^2\rangle} \right)\;,
\ee 
where $\langle \ldots\rangle$ indicate the time average.
This quantity has three drawbacks, namely i) it cannot be computed
without knowledge of the clean signal $\tilde{x}(t)$, ii) it can
be negative if the initial noise level is low, since distortion of
the signal by the filter can be stronger than the reduction of
noise, and iii) it does not quantify whether the intelligibility
of the signal was improved by the noise reduction. The latter is
easily seen by considering the limit of a very high noise level: in
such a case setting $y(t)=0\;\; \forall t$ will yield a positive gain,
even if this completely eliminates the signal.
Therefore, we
employ in this paper a commercial speech recognition program as a
quantifier of the success of noise reduction. The relative number
of words which are not correctly recognized is taken as a
quantifier for the signal corruption, regardless of whether this
is noise or some systematic distortion which might be introduced
by the noise reduction scheme. \par In this paper we briefly
recall the algorithm including its adaptation for the treatment of
voice data, which are nonstationary limit cycle-like signals with
embedded noisy segments (stemming from the fricatives). We then
apply it to data samples which without added noise are perfectly
recognized by the speech recognition software. We demonstrate that
our optimised noise reduction scheme does not reduce the
recognition rate when it is applied to clean data, and that it
improves the recognition rate when it is applied to noisy data,
which is comparable with a reduction of the noise amplitude by
about 1/2. Noise level $\%N$ is defined as 100 times the ratio of the
standard deviation of noise $\sigma$ and standard deviation of
data $\sigma_{data}$: $\%N=100 \cdot \sigma/\sigma_{data}$.

\begin{table}
\caption{\label{tab.1} The conceptional comparison between linear
methods, LP methods and Hybrid method.}
\begin{tabular}{||c||c|c|c|}
\hline \hline
    Methods/ & Linear methods & Nonlinear methods & Hybrid methods \\
    Concepts & (eg. low pass filter) & (eg. GHKSS) & (eg. LPNCF) \\
    \hline
    \hline
    Finding  & in  time& in space & in time \\
    neighbours & &  & and in space \\
    \hline
    Finding  & smoothing & smoothing & smoothing \\
    corrections & in time & in phase space  & using the information \\
     &  &   & in time and in space\\
    \hline
    Noise& in Fourier & violation  & violation \\
    estimation & space & of constraints & of constraints\\
     &  &  in phase space & in time and in space\\
    \hline
\end{tabular}
\end{table}
\section{Method}
For the purposes of noise reduction (NR) in voice signals we use
the LPNCF \cite{urbnrflow} and the GHKSS method \cite{GHKSS} for comparison. 
The GHKSS method is one version of the
Local Projective (LP)
\cite{Schreiber1,kostelich,Schreiber2,Farmer,hammel,davies,Hsu,Sauer,GHKSS}
method that was developed for chaotic signals corrupted by noise.
Assuming that the clean data are confined to some deterministic
attractor in a reconstructed state space, which itself is locally
a subset of a smooth manifold, the method aims at identifying this
local manifold in linear approximation and to project the noisy
state vector (which due to noise is not confined to this
hyperplane) onto the local manifold. Algorithmically, this means
to identify a neighbourhood in the delay embedding space and to
perform a Singular Value Decomposition of a particular covariance
matrix. Some refinements are described in~\cite{GHKSS}.
\par The LPNCF method, which was particularly developed for chaotic
flows, makes use of nonlinear constraints which appear because of
the time continuous character of the flow. These constraints are
functionals of the state vectors which assume the value $0$ for
dense sampling of deterministic data. Let $\set{x_i}$ for
$i=1,2,\ldots,N$ be the
 time series. We denote the corresponding clean
signal by $\set{\tilde{x}_i}$, so with the measurement
noise $\set{\eta_i}$ we have 
$x_i=\tilde{x}_i+\eta_i$ for $i=1,2,\ldots,N$. We define the
time delay vectors $\mathbf{x}_i=(x_i,x_{i-1},\ldots,x_{i-(d-1)})$
as our points in the reconstructed phase space. Then we choose
two neighbours $\mathbf{x}_k,\mathbf{x}_j\in\NN$ of the vector
$\mathbf{x}_n$ ($\NN$ is the set neighbours of the point
$\mathbf{x}_n$). Let us introduce the following function
\cite{urbnr03}
\begin{eqnarray} G_n(s)=x_{n-s}\lb x_{k+1-s}-x_{j+1-s}\rb\nonumber\\+x_{k-s} \lb
x_{j+1-s}-x_{n+1-s}\rb+x_{j-s}\lb
x_{n+1-s}-x_{k+1-s}\rb,\label{eq.g}\end{eqnarray} for
$s=0,1,\ldots,d-1$. \par The function $G_n(s)$ vanishes for clean
one-dimensional systems because it appears after eliminating $a$
and $b$ from following equations: \ber \tilde{x}_{n+1}=a\tilde{x}_n+b\nonumber\\
\tilde{x}_{k+1}=a\tilde{x}_k+b\nonumber\\
\tilde{x}_{j+1}=a\tilde{x}_j+b. \eer In the case of higher
dimensional systems the function $G_n(s)$ does not always vanish but
is altering slowly in time for dense sampling. \par Now one can
check that for a highly sampled clean dynamics there can be
derived such a constraint \begin{eqnarray} \mathbf{C}_n^q=\sum\limits_{k=0}^{q-1}(-1)^l
G_n(k)\approx 0,\nonumber\\ 
 l=\begin{cases}
    0 & if\; k=0, \\
    k+\sum_{s=1}^{\mbox{int}(\log_2(k))}
\mbox{int}(k/2^s) & if\; k>0,
  \end{cases}\label{eq.constraint}\end{eqnarray} where
$\mbox{int}(z)$ is a integer part of $z$ and $\log_2(z)$ is a
logarithm with a base $2$ of $z$. Similarly as in LP methods the
constraints~(\ref{eq.constraint}) are satisfied in this approach
by application of the method of Lagrange multipliers to an
appropriate cost function. Since we expect that corrections to
noisy data should be as small as possible, the cost function is
assumed to be the sum of squared corrections $S=\sum_{s=1}^N
\lb\delta x_s\rb^2$, where {$\delta x_n$} are the corrections of the NR method 
connected to {$x_n$}, such that the resulting time series of the 
NR method {$y_n$} 
is defined as $y_n=x_n+\delta x_n$ ($n=1,2,...,N$). The method is a
compromise between time and 
space integration methods. In the constraints there appear
neighbours in space together with their pre-images, 
and it works on a
time lag of unity in the embedding space in order to exploit the
flow-like structure of the data. 
Hence it combines spatial and temporal vicinity. It can perform better than
standard time averaging or standard LP methods, because 
the size of the neighbourhood in time and in space is smaller in
the LPNCF method than in standard methods which use only time or
space averaging. 
In a typical speech signal, only about 10-20 reasonable neighbours in
space exist, since a phoneme consists of about 10-20 slightly modified
repetitions of some basic pattern in time (see below and
Fig.~1). Therefore, an algorithm which tolerates a small number of
neighbours in space is required.
In this paper we use the LPNCF method as the main nonlinear method. 
The GHKSS method
is also employed for comparison but seems to be less
successful. Therefore, all technical details are only specified for
the LPNCF method in the following.

\begin{figure}
\includegraphics[scale=0.35,angle=-90]{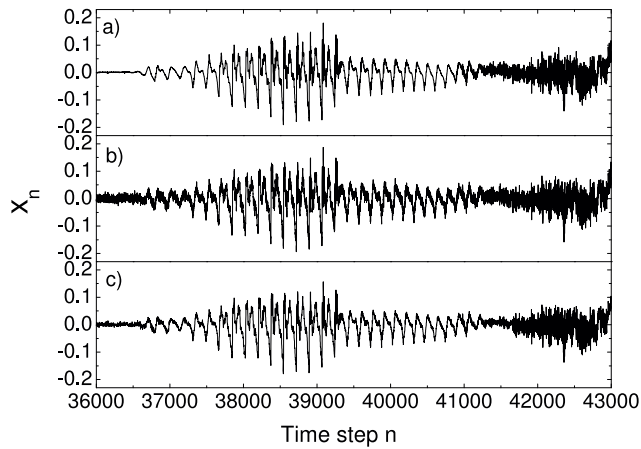}
\caption{\label{fig.signal} The voice time series of the word
"M\"{u}nchen", recorded with a sampling rate of 22050 Hz. 
The top panel is the clean signal,  
the next one shows the signal with
added noise ($\%N=35\%$), the bottom the signal after noise
reduction. The parameter $m$ calculated by the algorithm is as follows: 
for $n\in [35799,36501]$:  $m$=124, $n\in [36501,41122]$: $m$=168,
$n\in [41122,42716]$: $m$=7, $n>42716$: $m$=127.}
\end{figure}

It is known that the voice signal for most of the time has many
similarities with a flow \cite{kantzhuman}, which means it
represents smooth anharmonic oscillations with a typical frequency
around the speaker's pitch of a few hundred Hz.
However, articulated human voice is a
concatenation of different phonemes, so that the frequency,
amplitude and, most importantly, the wave form of the oscillation
various tremendously on time scales of about 50 to 200 ms, causing
the signal to be highly nonstationary. A qualitatively
different component in articulated human voice is due to
fricatives and sibilants, which are high frequency broad band
noise-like parts of the signal. Such a sound starts around $n$=41200
in Fig.~\ref{fig.signal}. All nonlinear noise reduction
schemes are very suitable for removing noise from anharmonic
oscillations but they have the tendency to suppress strongly the
fricatives and sibilants. Since the latter are of utmost relevance
for a correct recognition by a speech recognition algorithm, we
have to take special care of these. Finally, there are pauses in
the speech which are pure noise after noise contamination of the
signal. It is important to remove the noise during the speech
pauses, so that the beginning and ending of words is correctly
identified by the recognition algorithm. A particular challenge
lies in these two opposing requirements: noise like fricatives
should not be suppressed, whereas noise during speech pauses must
be eliminated.
\par So the important modification of the standard algorithms for
stationary data here is to identify the fricatives/sibilants and
to treat them in a different way. As a first step we compute the
auto-correlation function in a gliding window analysis (using
windows of 300 sample points, which is about 14 ms). The location
of the first maximum serves as a rough estimate of the dominant
period in the signal. We can then define windows in time during
which the dominant frequency is nearly constant. Obtaining the
autocorrelation function is rather fast because we use previous
calculations in next windows.
\par The estimated period inside a window is
used to fix almost all of the parameters of the algorithm, e.g.,
the embedding dimension and embedding delay in nearest
neighbourhood searching, the maximal embedding projection, a
maximal range of neighbourhood in time etc. We also optimize some
initial low-pass filtering of the signal by simple averaging on
time windows which are about 1/40 of the local period determined
by the auto-correlation function. The most important parameters of 
the algorithm are the embedding dimension in neighbourhood searching
$m$, the minimal and maximal number of nearest 
neighbours, and the  embedding dimension $q$ on 
which the noisy delay vectors are projected.
Due to the nonstationarity of human speech, the total time window
covered by a delay vector should be identical to the estimated
period. Only this guarantees that the wave forms of different
phonemes can be fully distinguished from each other and correct neighbours are
identified. For the LPNCF method the embedding dimension $m$ should be
 equal to the period with a time
lag of unity, whereas in the GHKSS
algorithm one would use an embedding window of the same size, but
reduce the embedding dimension by the introduction of a time lag
larger than unity. The maximal number of nearest 
neighbours was fixed to 12 here, the dimension $q$ of the submanifold
for projections was varied in between
4 and 24
for the LPNCF method and between 3 and 12 for GHKSS method. For this
latter parameter, there exist ways to adapt it automatically to the
dynamics \cite{GHKSS,greeks}.
\par
The way that we control the parameters of the noise
reduction method through the observed period is such that for large
periods the outcoming signal after reduction is much smoother than
for short periods. This means that large corrections are made
on those parts of the signal where large periods are detected.
If we detect a short autocorrelation time and a low variance, both
less than some threshold,
we consider this part of the signal as pure noise inside a speech
pause. 
We hence overwrite the computed period and set it to 
its maximal value, in order to
 flatten the signal to zero. Sounds like 's','tch','h' are like noise
 with very little periodicity but 
the energy flow (here variance) is much higher than for the noise
(e.g., in Fig.~\ref{fig.signal} at $n$=41200 begins
'chen'). In order to prevent the algorithm from removing these
parts of the signal, we do not do any corrections when the
detected period is less than 6 sample intervals and the variance
is high. All parameters were optimized empirically and might
depend on the voice recognition software, on the language spoken,
and perhaps even on the speaker.




\section{\label{sec.linear}Linear filter}
\par In order to do a comparison to the well-known linear
filtering we apply the Wiener filter to the noisy signals. Since
we use white noise, the noise spectrum is fully determined by the
noise level $\sigma$. Knowing the noise level we can employ a perfect
linear filter, but in practice we have to estimate the noise level
from
 the data using the power spectrum. For the purposes of our
linear filter we estimate the noise standard deviation $\sigma$
using the upper $40\%$ percent of the frequency domain, where
the power spectrum is flat. Then the Wiener filter can be
described as follows. If $S_k^{signal}$ is the amplitude of
Fourier Transform of the noisy signal, additivity of the noise and
independence of the noise and signal tell us: \be
(S_k^{signal})^2=(S_k^{noise})^2+(S_k^{clean})^2.\ee The action of
the optimal linear filter for white noise consists in rescaling
the amplitudes in Fourier space of the signal by the use of noise
variance: 
\be S_k^{after}=S_k^{signal}\left(
\frac{(S_k^{signal})^2-\sigma^2}{(S_k^{signal})^2}\right).\ee 
The inverse Fourier transform on the corrected amplitudes
$S_k^{after}$ keeping the phases of the Fourier transform of the noisy
signal yields the new signal. One can prove that knowing the exact
value of the noise level $\sigma$
such an algorithm is the optimal linear method of noise reduction
\cite{NR}.

\section{Results}
The speech recognition is done by the commercial software program
Linguatec ViaVoice Pro release 10 for German, which is based on
the IBM recognition algorithm viavoice. The difficulty in speech
recognition lies in the required training of the algorithm in
order to adapt to a specific speaker. In order to make our results
reproducible, we downloaded the sample sentences together with the
speaker specific auxiliary data files, from the
distributor\cite{Linguatec}.
\par We were working on the following recorded sentences in German: \\
"M\"{u}nchen, der 21.10.04. Sehr geehrter Herr Schneider, Sicher
werden wir noch viel zu besprechen haben. Das Problem liegt
offensichtlich an der Funktionsvielfalt. Ein Vertragsabschluss
kann von uns nur erfolgen, wenn auch eine Konventionalstrafe
vereinbart wird, und zwar in H\"{o}he von € 1.000,- pro Tag
Verzug. Dies wird voraussichtlich in der ersten Juniwoche sein.
Mit freundlichen Gr\"{u}{\ss}en".
\par 


In order to create noisy signals, we first convert the
speech stored in the {\tt wave}-format (.wav) into real numbers,
representing the time series of the sound amplitude. We add
independently drawn Gaussian random numbers of the desired
variance and apply the back-conversion into the {\tt wave}-format
for the determination of the recognition rate. The noise reduction
is done on the real valued time series, again with a successive
conversion for recognition. In Fig.~\ref{fig.signal} we show the
signal which corresponds to the word "M\"{u}nchen". In the upper
panel (a) there is a clean signal. In the middle, part (b), the
noisy signal with standard deviation (SD) of noise equal 0.009 and
in the bottom, part (c), the noisy signal after NR. As pointed out
before, around n=41200 is the fricative ``ch'' (pronounced as
[\c{c}]). The autocorrelation function suggests a period of 4, and
the variance is much larger than for noise on a pause which can be
seen on the beginning of the signal (b). The signal (a) and (c)
are recognized well by the program but the noisy signal (b) is
badly represented in the recognition of the full text. Hence,
although the noise level appears to be small, the recognition
software is considerably confused and the recognition rate drops
significantly.
\par
\begin{figure}
\includegraphics[scale=1]{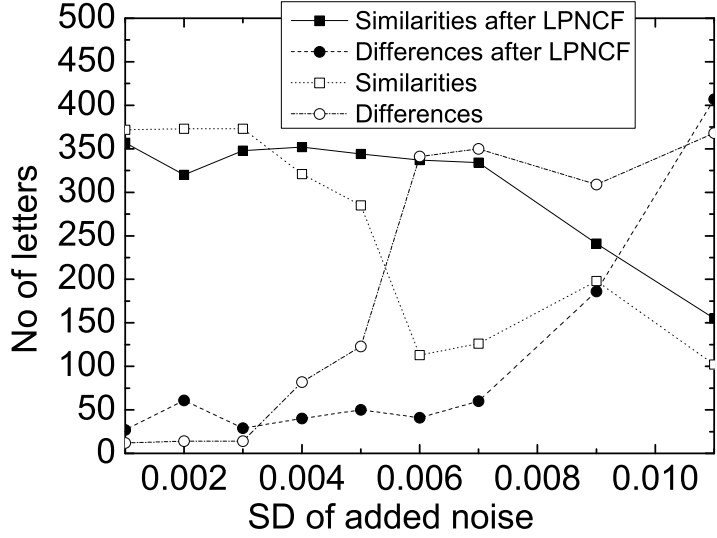}
\caption{\label{fig.simdiff} Plot of similarities and differences
in letters of the correct text and texts recognized from noisy
signals (open squares and circles) or texts recognized from noisy
signals after nonlinear noise reduction (solid squares and circles). Standard
deviation of added noise appears on the x-axis.}
\end{figure}
\begin{figure}
\includegraphics[scale=1]{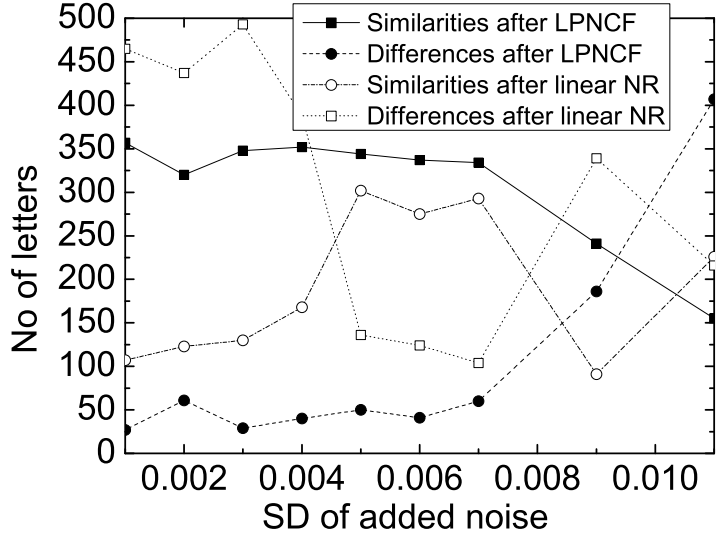}
\caption{\label{fig.simdifflin} Plot of similarities and
differences in letters of the correct text and texts recognized from noisy signals
after noise reduction using LPNCF method (solid squares and circles) or texts recognized from
noisy signals after optimal linear noise reduction (open
squares and circles). Standard deviation of added noise appears at the
x-axis.}
\end{figure}
\begin{figure}
\includegraphics[scale=1]{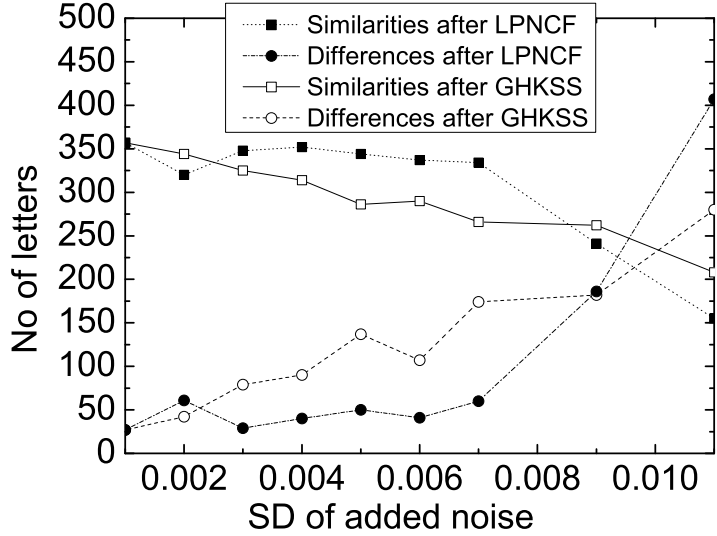}
\caption{\label{fig.simdiffghkss} Plot of similarities and
differences in letters of the correct text and texts recognized from noise signals
after noise reduction using LPNCF method (solid squares and circles) or texts recognized from
noisy signals after noise reduction using GHKSS method (open squares and circles). Standard deviation of added noise appears at the
x-axis.}
\end{figure}
\begin{figure}
\includegraphics[scale=1]{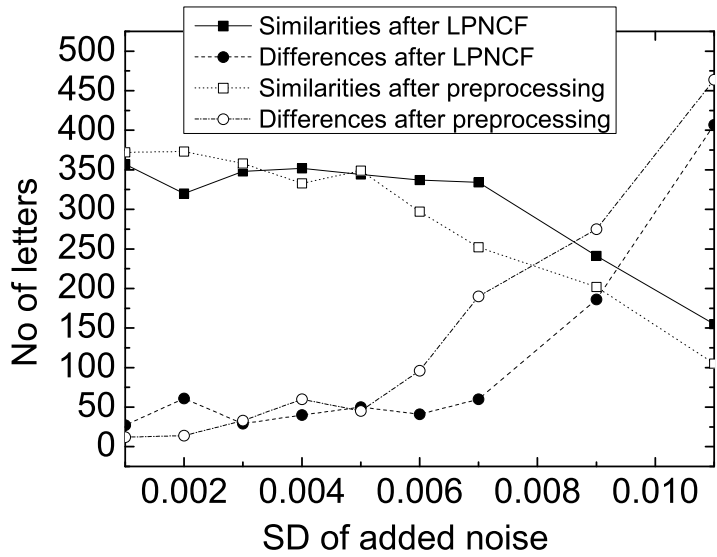}
\caption{\label{fig.simdiffav} Plot of similarities and
differences in letters of the correct text and texts recognized from noisy signals
after noise reduction using LPNCF method (solid squares and circles) or texts recognized from
noisy signals after preprocessing analysis (open squares and circles). Standard deviation of added noise appears at the
x-axis.}
\end{figure}

The algorithm of the recognition program enforces it to generate
reasonable words only, which is, it only generates words form an
internal dictionary. Therefore, misunderstanding by the
recognition software cannot lead to wrong letters inside words,
but only to the replacement of correct words by some other words.
Rarely, the wrongly recognised word resembles in sound the
original word. If the system is strongly misled, it can generate a
long wrong word out of several short ones or vice versa, such that
the number of words is not conserved. However, the total number of
letters is more or less unchanged. Hence, in order to do
statistics on the recognised sentences, we use the following
indicator: We identify the correctly recognised words and those
words which are not part of the original sentences, and then count
the numbers of letters inside these two groups of words.
In Figs~\ref{fig.simdifflin}~-~\ref{fig.simdiffav}
we show these differences and
similarities as a function of the amount of noise added. Without
noise reduction 
(see Fig.~\ref{fig.simdiff}),
a standard deviation of the added noise of more
than 0.003 leads to mis-interpretations of the speech recognition
software. If one takes into account that every wrongly recognised
letter requires a correction by hand, the recognition is useless
when more than half of the number of characters has to be
replaced. This situation occurs for noise levels above 
0.005 ($\%N=20\%$). 
After noise reduction, the recognition rate
increases considerably. However, for very low noises distortions
of the signal introduced by the noise filtering leads to a small
number of wrongly recognised letters.



\par For comparison the results of the best linear filter which was described in
section~\ref{sec.linear} are shown in Fig.~\ref{fig.simdifflin}.
 We see that such a filter reduces the recognition rate for small noise levels and improves for high ones.
For low noise levels, these are overestimated from the power
spectrum, so that the high-frequency components of the signal are
strongly reduced. In the case of high noise levels their
estimation is rather good so reduction is well done improving the
recognition ability.
\par

In order to do some comprehensive comparison to standard local
projective methods we also employed the GHKSS noise reduction method (see
Fig.~\ref{fig.simdiffghkss}). The LPNCF method works better for
intermediate noise levels in the sense that the resulting signals are
more useful for the speech 
recognition program. Also for comparison, in Fig.~\ref{fig.simdiffav} 
we present the recognition rate after pure preprocessing of the data
in the above described way by detecting the period and performing a
period depending low-pass. This shows that the detection of the
correct beginnings and endings of words, which is improved by the
preprocessing, as well as the simple averaging
implies an improvement of recognition, but that the nonlinear filter
on top of that also contributes to the success.


\par
\begin{figure}
\includegraphics[scale=1]{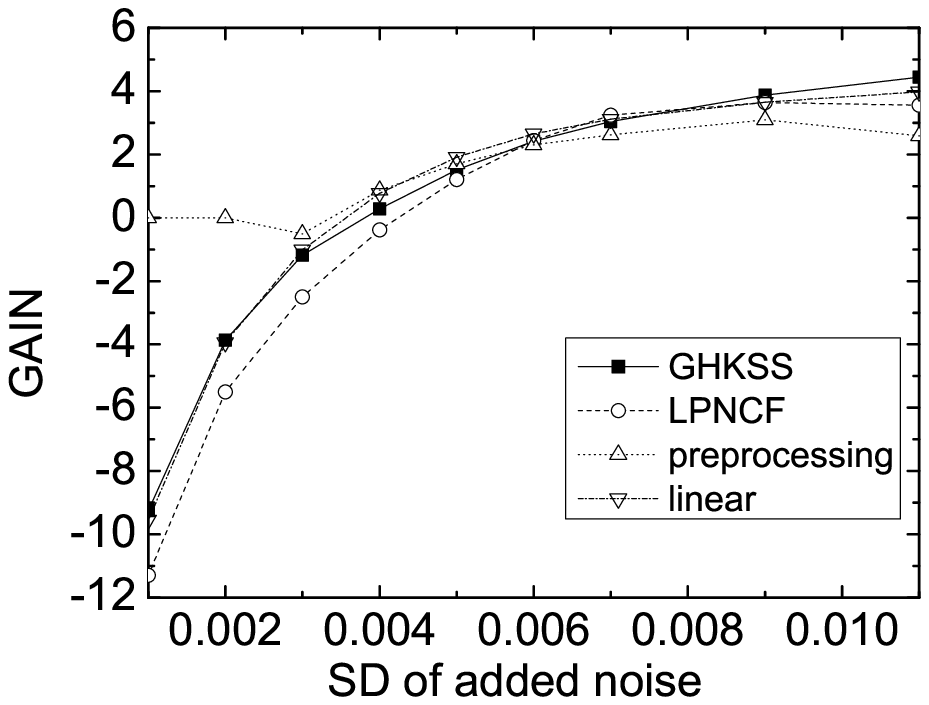}
\caption{\label{fig.gain} Plot of the gain parameter versus SD of
added noise in the signal.}
\end{figure}

The gain parameter corresponding to
Figs~\ref{fig.simdifflin}~-~\ref{fig.simdiffav} 
is presented in
Fig.~\ref{fig.gain}. One can see that the efficiency of nonlinear
and linear noise reduction is comparable in these both cases and
not very high especially for small noise levels. 
The pure preprocessing does not lead to any correction at all in the
low noise limit. However, this is a desirable feature, since the other
methods even have a negative gain for
low noise levels. The almost clean signal is more
distorted by either of the noise reduction schemes than there is noise
to be removed. The nonlinear
methods introduce some distortion everywhere where the voice is
not well represented by a flow and is not sufficiently smooth (see
Fig.~\ref{fig.signal}). Also for larger noise levels the gain is
small compared to gains obtained in \cite{IEEE}, which reflects
that the data structures which must be preserved for a good
recognition cannot be directly translated into gain. Also,
however, the noise levels which are relevant in the present study
are much smaller than those considered in \cite{IEEE}, since
larger noise levels completely destroy the speech recognition.
 We see that the gain parameter is not a good indicator of the
 recognition rate. This property is very pronounced for small noise
 levels, as a comparison of the recognition rates of 
nonlinear and linear filters shows. The similar gains lead to 
recognition rates in favour of nonlinear NR. Even if the gain is
negative the program for speech recognition is not much mislead in the
nonlinear NR case. 

\section{Conclusions}
Due to the specific
 properties of articulated human voice with strong
non-stationarity, but also with an interplay of clear phonemes and
noise-like sounds such as fricatives, noise reduction of human speech
signals is not a straight-forward task. Many specific properties can
be taken into account in order to improve any chosen algorithm. In
this paper we employ a variant of the class of nonlinear noise
reduction schemes, the LPNCF method, together with some adjustments
which take the specific properties of speech into account. Mainly,
this is a detection scheme which distinguishes between noisy speech
pauses and noise-like parts of voice, the fricatives and
sibilants. With this modification, we achieve considerable
improvements of the recognition rate of an automatic commercial speech
recognition program, which corresponds to roughly reducing the noise
amplitude by a factor of two. 

The whole scheme works in the low noise regime only, since the voice
recognition program fails already for noise amplitudes which for the
human ear appear rather low. For the optimization of parameters of the
various schemes,
the noise free signal is not needed. However, what we need is the
correct meaning of the spoken text in order to measure the recognition
rate and thereby optimizing the performance of the methods. In
practice, the latter is not a strong restriction, since due to the
smallness of the admissible noises, every human will correctly
understand the spoken sentences. 

Future work will be of more technical nature, whereas this article was
aiming to demonstrate that 
noise removing using chaos like features improves the recognition
rate especially for intermediate noise levels and does not destroy
the signal when the noise level is small. First of all, the generality
of the results and in particular of the parameter settings has to be
tested on recordings from various speakers. Secondly, further
improvement of the program speed might be useful, although with
current cycle times of PCs the LPNCF method is already close to real time.
In the long run, we plan
to implement this algorithm in a microprocessor in order to do on-line
preprocessing of the microphone input to the computer on which the
speech recognition software is running.

\section{acknowledgement}
The work has been supported by the Project  STOCHDYN by European Science
Foundation and  by Polish  Ministry of  Science and Higher Education (Grant
15/ESF/2006/03).
\newpage

\end{document}